# Structural reconstruction and anisotropic conductance in 4*f*-ferromagnetic monolayer

Haipeng You [a,#], Jun Chen [a,#], Jun-Jie Zhang [b,#], Ning Ding [a], Xiwen Zhang [a], Xiaoyan Yao [a,*],

Shuai Dong [a,†]

[a]*School of Physics, Southeast University, Nanjing 211189, China*

[b]*Department of Materials Science and Nano-Engineering, Rice University, Houston, Texas 77005, USA*

A B S T R A C T

Two-dimensional magnets are promising for nanoscale spintronic applications. Currently, most available candidates are based on 3*d* transition metal compounds, with hexagonal or honeycomb lattice geometry. Here, a $GdCl_3$ monolayer with 4*f* moments is theoretically studied, which can be exfoliated from its existing bulk. Its orthorhombic structure and hendecahedral ion cages are unique in two-dimensional. Furthermore, a significant structural reconstruction is caused by the implantation of Li atoms into its interstitial position, which also lead to ferromagnetism via a double-exchange-like process. Its highly anisotropic conductance may be peculiarly useful for nanoelectronics.

[†]Corresponding Authors

Email address: sdong@seu.edu.cn

[*]Corresponding Authors

Email address: yaoxiaoyan@seu.edu.cn

# These authors contributed equally

# 1. Introduction

The discovery of intrinsic ferromagnetism in two-dimensional (2D) materials, such as $CrI_3$ monolayer and $Cr_2Ge_2Te_6$ few layers [1,2], brings promising possibilities of further miniaturization and high integratability for spintronic devices [3-7]. Nonetheless, their immediate applications are limited by low Curie temperatures ($T_C$'s) and other drawbacks such as material instabilities. Hence, the exploration of more 2D robust magnets with better physical properties and more functionalities, remains the most vital scientific issue in this field, which is crucial for both fundamental science and applications.

It is noteworthy that most known 2D magnets, no matter experimentally verified or theoretical predicted, are based on spin moments of $3d$ electrons, whose orbitals are spatially localized with strong Hubbard correlation [8-21]. Apart from these transition metal compounds with unpaired $3d$ electrons, rare earth compounds with unpaired $4f$ electrons also play a significant role in modern magnets. Comparing with $3d$ electrons, $4f$ electrons own larger upper-limit of moments, stronger Hubbard correlation, and stronger spin-orbit coupling (SOC). These characteristics can be utilized in 2D materials to pursuit nontrivial magnetism.

Recently, Wang *et al.* predicted a 2D hexagonal ferromagnetic (FM) monolayer $GdI_2$ with a high Curie temperature ($T_C$) 241 K and a large magnetic moment 8 $\mu_B$/f.u. [22], which opened the door to pursue prominent 2D $4f$-magnets. However, its magnetic easy-plane is disadvantage to stabilize the long-range FM order, according to the Mermin-Wagner theory. Later, our work studied another gadolinium chloride ($GdI_3$) monolayer with honeycomb geometry [23]. Its

Néel-type antiferromagnetic (AFM) ground state can be changed to a stripy-AFM order upon the Li or Mg intercalations in its interstitial positions of hexatomic rings. Peierls transition occurs due to the strong electron-phonon coupling, which induces ferroelasticity in $(GdI_3)_2Li$ and $(GdI_3)_2Mg$ [23].

In this work, one more 2D gadolinium halide $GdCl_3$ will be studied using density functional theory (DFT). Noteworthy, it is not a marginal extension of aforementioned sister compounds but owns conceptually distinct physics. First, different from neither the hexagonal structure of $GdI_2$ nor honeycomb structure of $GdI_3$ which are common in 2D materials [24], the $GdCl_3$ monolayer owns an unusual orthorhombic geometry, where each Gd ion is caged within a rare hendecahedron. In this sense, $GdCl_3$ represents a brand-new lattice for 2D magnets. Second, the doping of Li atoms into its interstitial positions will cause a significant structural reconstruction. Consequently, its magnetic ground state changes from the AFM to FM one. Last but not the least, distinctive conductive anisotropy is found, which may be naturally useful for nanoelectronic cabling. These peculiar physical properties and exotic scientific mechanisms certainly go beyond previous studies, and open a much broader scope of 2D magnets.

## 2. Methods

First-principles calculations are performed based on spin-polarized density functional theory, which is implemented in Vienna ab initio Simulation Package (VASP) with projector-augmented wave potentials [25]. The generalized gradient approximation with Perdew-Burke-Ernzerhof is used for the exchange-correlation functional [26]. The kinetic energy cutoff for the plane wave is set as 500 eV. The standard pseudopotentials are used, and the valence electron configuration is treated as $5s^25p^64f^75d^16s^2$ for Gd. The Hubbard correlation is considered by using GGA+$U$ method with $U$=9.2 eV and $J$=1.2 eV applied on Gd's 4$f$ orbitals [27]. Spin-orbit coupling effect is used to calculate its magnetocrystalline anisotropy energy.

The total energy convergence criterion of $10^{-6}$ eV and force convergence criterion of $10^{-2}$ eV/Å are chosen during the structural relaxations. To accommodate possible magnetic states, a 2×2×1 supercell is used. A 7×7×1 k-grid is adopted to sample the Brillouin zone by utilizing the Monkhorsto-Pack k-points scheme for the unit cell [28]. The van der Waals (vdW) correction is considered for its bulk by using the DFT-D2 method [29]. A vacuum space of 20 Å thickness along the out-of-plane direction is adopted to eliminate the interaction between adjacent layers.

To account the finite-temperature effects, the Markov-chain Monte Carlo (MC) method with Metropolis algorithm is used to simulate the magnetic phase transition [30], with magnetic coefficients extracted from DFT energies. A 24×24 orthorhombic lattice with period boundary conditions is used. Parallel tempering algorithm is used to obtain the thermal equilibrium efficiently [31], with an exchange sampling after every 10 standard MC steps. Typically, the initial $3×10^4$ steps are discarded for the equilibrium consideration and additional $3×10^4$ steps are retained for the statistic averaging of simulation. We also check the results simulated on a larger lattice of up to 36×36 to ensure that the finite-size effect would not change our conclusion.

In addition, the *ab initio* molecular dynamics (AIMD) simulation in the *NVT* ensemble lasts for 6 ps with a time step of 1 fs. The temperature is controlled by using the Nose-Hoover method [32].

The electrical conductivity is calculated by solving the semi-classical Boltzmann transport equations based on constant relaxation-time approximation (here =10 fs) implemented in the BoltzWann code [33]. This approach employs a generalized Wannier-Fourier interpolation technique [34] in order to obtain electron eigenvalues on dense Brillouin zone (BZ) grids by means of maximally localized Wannier functions (MLWFs) [35]. Here, the 2D BZ of 7×3 grid *k*-points is used to construct the MLWFs, while the dense BZ (180×120 grid *k*-points) is utilized to calculate the electrical conductivity.

## 3. Results and discussion

Four possible structures of GdCl$_3$ bulk were reported according to the Materials Project Database [36] and early experiment [37], as shown in Fig. 1. Among them, the *P*-phases (*P*6$_3$/*mmc* and *P*6$_3$/*m*) are constructed by one-dimensional (1D) chains coupled via van der Waals (vdW) interaction. The *R*-phase (*R-3*) possesses the vdW layered structure with ABC-type stacking, which is usual for many other *MX*$_3$ compounds like GdI$_3$ [23, 24]. The unusual *Cmcm*-phase with PuBr$_3$-type vdW stacking in the AB sequence, was reported in the 1960s [37].

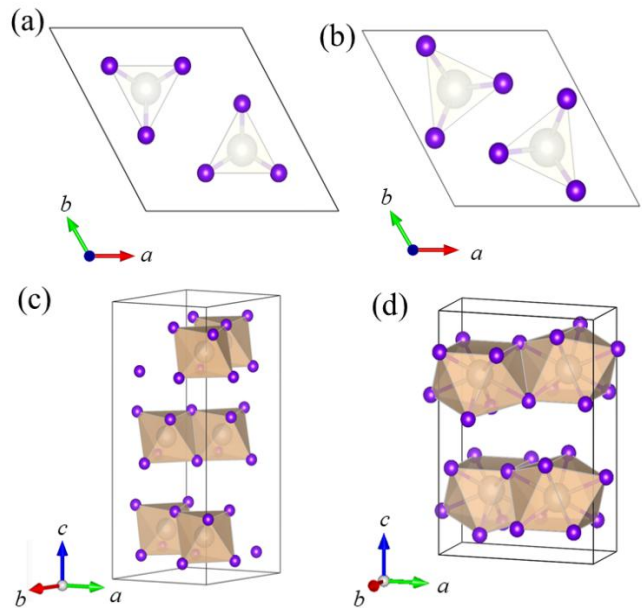

**Fig. 1.** Possible crystal structures of GdCl$_3$ bulk. Gray: Gd; purple: Cl. (a-b) The quasi-1D *P*-phases with space groups *P*6$_3$/*mmc* and *P*6$_3$/*m*. (c) The *R-3* phase. (d) The *Cmcm* phase.

**Table 1.** Comparison of GdCl$_3$ bulks and monolayer. Lattice constants (in unit of Å), energies (in unit of meV/f.u.), magnetic moments (in unit of $\mu_B$/Gd), and band gaps (in unit of eV) are from DFT calculations. The energy of *Cmcm* bulk is taken as the reference, with unified ferromagnetic (FM) order for all candidates. The experimental (Exp) one is shown for comparison [37].

| GdCl$_3$ | Space group | a | b | c | Energy | Moment | Gap |
|---|---|---|---|---|---|---|---|
| bulk | *P6$_3$/mmc* | 9.022 | 9.022 | 3.650 | 406 | 7.082 | 3.1 |
| bulk | *R-3* | 6.924 | 6.924 | 17.875 | 107 | 7.087 | 4.6 |
| bulk | *P6$_3$/m* | 7.353 | 7.353 | 4.100 | 22 | 7.062 | 4.1 |
| bulk | *Cmcm* | 3.856 | 8.544 | 11.849 | 0 | 7.070 | 3.3 |
| bulk (Exp) | *Cmcm* | 3.88 | 8.52 | 11.73 | - | - | - |
| monolayer | *Pmmn* | 3.854 | 8.615 | - | - | 7.071 | 3.9 |

All these possible structures are calculated for comparison. As compared in Table 1, the *Cmcm* phase owns the lowest energy, implying the ground phase, consistent with the experiment [37]. The optimized lattice constants of *Cmcm* phase also agree well with the experimental ones [37], proving the reliability of our calculation. The magnetic moment is ~7 $\mu_B$/Gd, originating from its half-filling 4*f* orbitals.

Starting from these reliable results of bulk, the GdCl$_3$ monolayer is investigated. The GdCl$_3$ monolayer (with space group *Pmmn*) can be easily exfoliated from its vdW layered bulk, because its cleavage energy (0.25 J/m$^2$) is smaller than that of graphite (0.37 J/m$^2$), as shown in Supplementary Fig. S1. Due to the smaller ionic radius and stronger electronegativity of Cl, the Gd-Cl bond lengths (2.67-2.95 Å here) are shorter than the Cd-I one (3.07 Å) in GdI$_3$ [23]. As a result, the ion cage wrapping each Gd is more compact in GdCl$_3$, forming a rare hendecahedron with two more ions than common octahedron or triangular prism. Neighboring hendecahedra connect with each other by sharing a face of three Cl ions along the *a*-axis, and an edge of two Cl ions along the *b*-axis, as shown in Fig. 2a.

To study its magnetism, four most possible magnetic configurations are considered, including FM and three AFM ones: Stripy, Zigzag, and Double-Stripy (DS), as shown in Fig. 2a-d. Among these candidates, the Zigzag-AFM phase is found to own the lowest energy, as compared in Supplementary Table S1. The physical origin is that the half-filling 4*f* orbitals prefer AFM exchanges according to the Goodenough-Kanamori rule. And the shorter Gd-Cl$_3$-Gd bonds along the *a*-axis lead to stronger exchanges comparing with the longer Gd-Cl$_2$-bonds, and thus dominate the AFM texture in these isosceles triangles. The electronic structure of Zigzag-AFM GdCl$_3$ monolayer is shown in Fig. 2e, demonstrating a Mott insulator with a band gap ~3.9 eV. The valance band maximum originates from the 3*p* orbitals of Cl and the conduction band minimum is contributed by Gd's 5*d* orbitals.

Since Gd's 4*f* orbitals are very localized and far below the Fermi level [see Fig. 2e], the magnetic exchanges will be consequentially weak, consistent with the small energy differences between FM and various AFM phases. Besides the exchanges, magnetic anisotropic energy (MAE) is vital to stabilize a long-range magnetic order in the 2D limit. According to our DFT calculation with SOC, the easy axis of GdCl$_3$ monolayer is pointing along the *a*-axis, but only 0.12 meV/Gd and 0.06 meV/Gd lower than the cases pointing along the *b*-axis and *c*-axis respectively. Such a weak MAE is reasonable, since the orbital moment of half-filling 4*f* orbitals is mostly quenched and thus the effective SOC effect is seriously suppressed. Thus, its Néel temperature should be very low, common for pure 4*f* magnets.

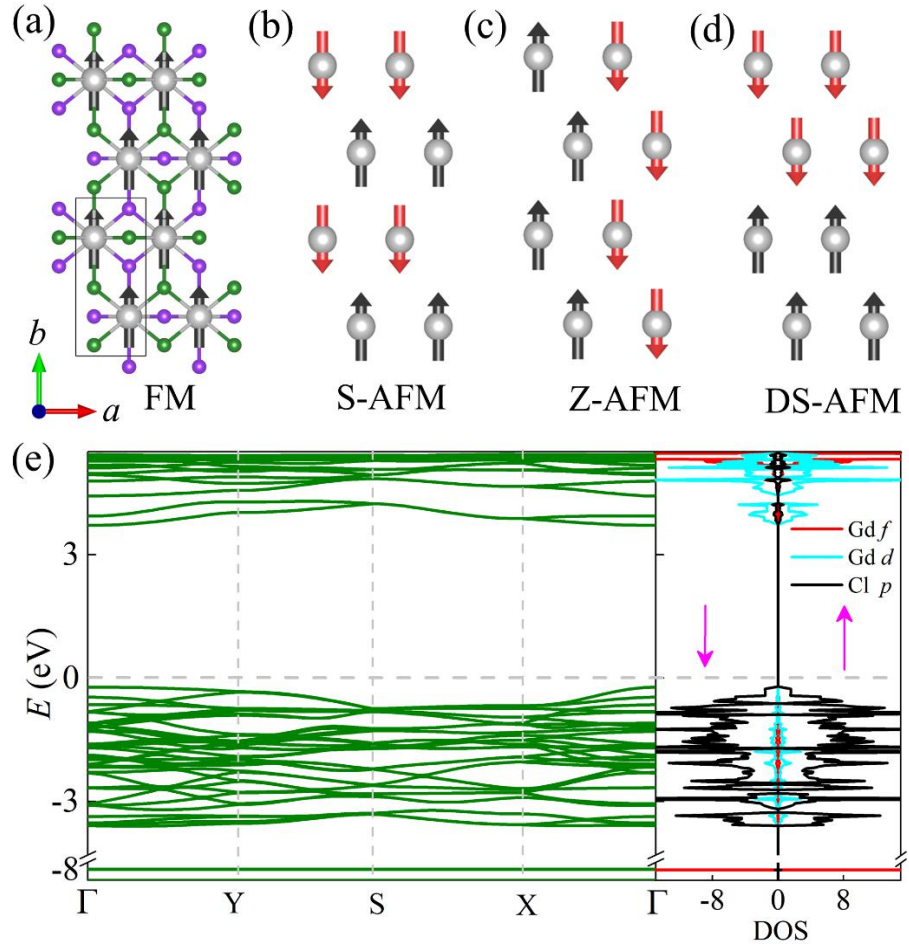

**Fig. 2.** Four possible magnetic orders of GdCl$_3$ monolayer. The upper and lower Cl ions are distinguished by colors. (a) FM. (b) Stripy-AFM. (c) Zigzag-AFM. (d) DS-AFM. The primitive cell is indicated by the rectangle in (a). In (b-d), all Cl ions are omitted for simplicity. (e) The band structure and atom-projected density of states (DOS) of the Zigzag-AFM order. The very narrow 4*f* electronic bands appear at ~-8 eV.

Aforementioned results seem to suggest GdCl$_3$ monolayer a low-value 2D material with low-temperature antiferromagnetism, even if its structure is somewhat exotic. It may be the reason why it is being ignored even if its parent bulk has been synthesized for more than half century [37]. Even though, the variable valence of Gd provides the feasibility to tune its magnetism via electron doping.

Here the doping is mimicked by placing one Li atom per u.c. in different initial positions, and then relaxing these structures. The possible configurations are indicated as the α-, β-, and γ-states shown in Fig. 3a. Now, the chemical formula of monolayer becomes $(GdCl_3)_2Li$. According to our DFT calculation, the γ-state, with Li implanted in the largest interstitial position between upper and lower Cl ions, owns the lowest energy. Although the original interstitial vacancy (diameter ~ 3.32 Å) seems enough to accommodate a $Li^+$ ion (diameter ~ 1.52 Å), unexpectedly, the γ-state of $(GdCl_3)_2Li$ monolayer undergoes a significant structural reconstruction during the relaxation. After this Li implantation, the exotic Cl-hendecahedron cages spontaneously transform into the usual Cl-octahedron cages, wrapping both Gd and Li ions, as shown in Fig. 3b-c. The symmetry of monolayer is then reduced to monoclinic ($C2/m$). The AIMD simulation at 300 K further confirms its thermal stability after the structural reconstruction (Fig. S2). The geometric distortion of Gd's isosceles triangles can be viewed in Fig. 3(d).

For comparison, as shown in Supplementary Fig. S3, the the optimized α-state does not have significant structural reconstruction, and Li atoms only slightly move vertically downward. However, there is also obvious structural reconstruction for the optimized β-state. Even though, the reconstructed β-state is higher in energy than that of γ-state. Thus, only the γ-state phase is presented in this work since it possesses lowest energy, which is the most critical result for (GdCl$_3$)$_2$Li monolayer.

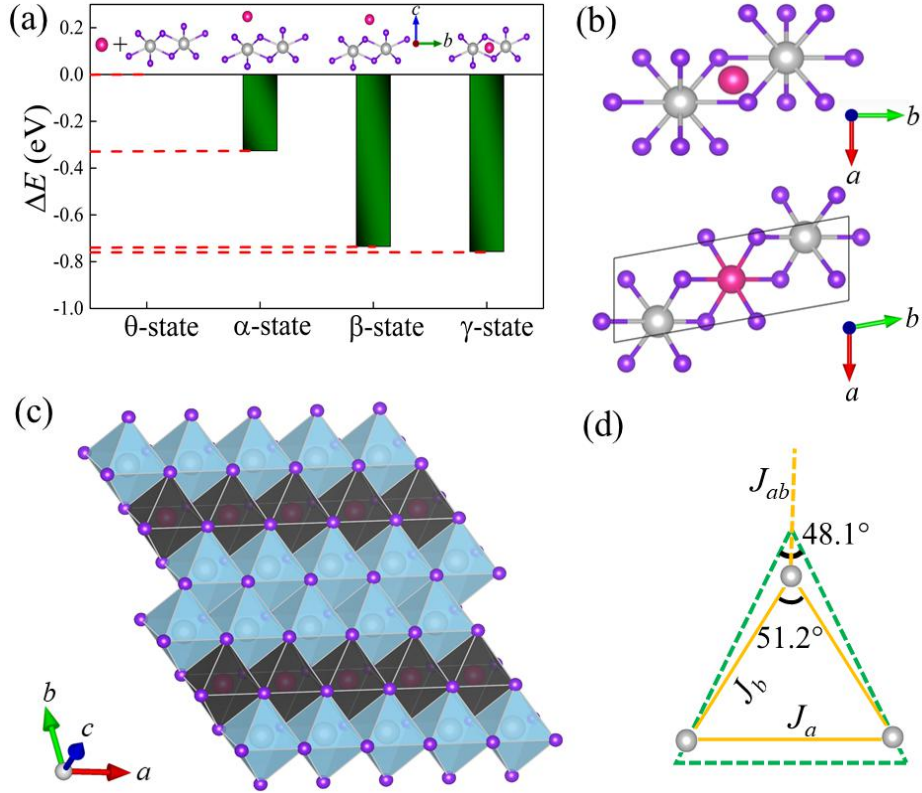

**Fig. 3.** The structural reconstruction caused by Li doping to GdCl$_3$ monolayer. Red: Li. (a) The energies for different initial positions of Li (denoted as α-, β-, and γ-states) after the optimization. The energy of θ-state (isolated Li atom and GdCl$_3$ monolayer) is taken as the reference. (b) The structures of γ-state before and after the structural optimization. The primitive cell is indicated by the parallelogram. (c) The stereoscopic view of (GdCl$_3$)$_2$Li with edge-sharing Cl-octahedra. The Gd's bi-stripes are separated by Li chains. (d) Comparison of Gd's isosceles triangles. Green: GdCl$_3$; Yellow: bi-stripy in (GdCl$_3$)$_2$Li. The triangles are shrunk in the doped case: -9.6%/-4.2% for the long/short side respectively, although the apical angle is slightly enlarged. The intra-stripy exchanges are indicated as $J_a$, and $J_b$, while $J_{ab}$ denotes the inter-stripy one.

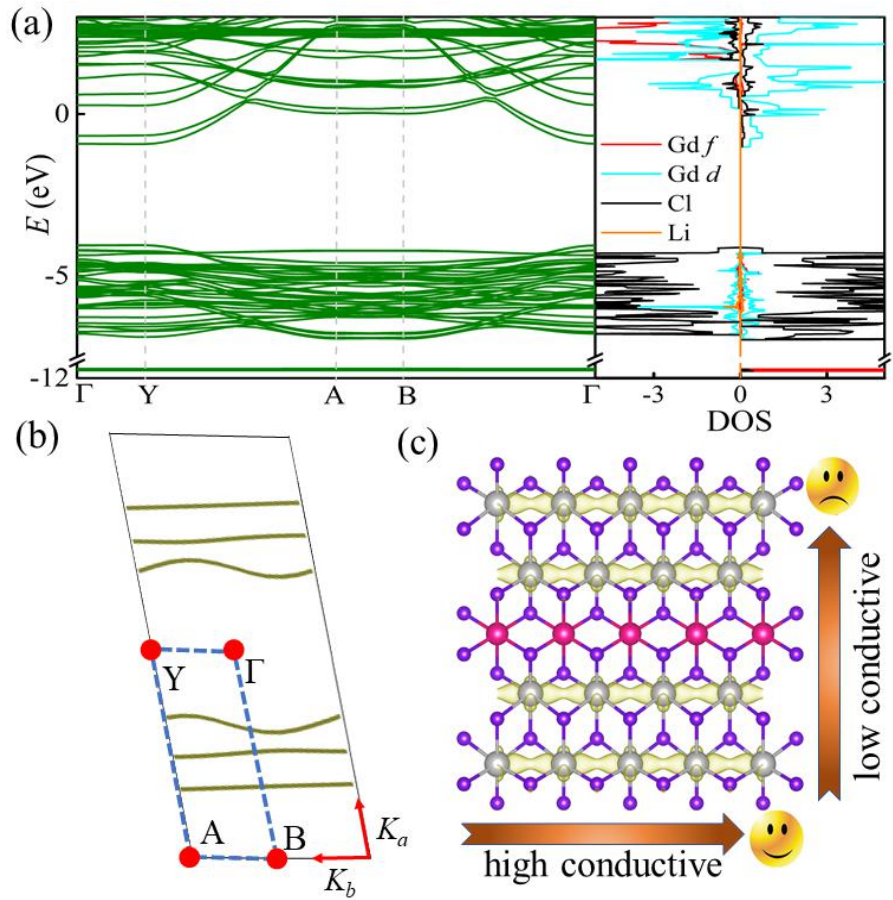

**Fig. 4.** Electronic structure of $(GdCl_3)_2Li$ monolayer. (a) The band structure and atom-projected DOS. The Fermi level is shifted upward to cross the $5d$ bands of Gd. (b) The Fermi surfaces in the 1st Brillouin zone, which consist of six 1D curves. (c) The electronic density distribution, which prefers the high conductance and direct exchange between Gd's spins along the $a$-axis.

After the doping, the possible magnetic states are re-calculated, as summarized in Supplementary Table S2 of SM. Now the FM state becomes the lowest energy one, which is desired for applications. The DS-AFM state, with FM order within each bi-stripy but AFM order between neighboring bi-stripes, owns a slightly higher energy, implying a very weak correlation between neighboring bi-stripes. In other words, the ferromagnetism in $(GdCl_3)_2Li$ is quasi-1D, due to the separation of Li's chains [see Fig. 3c]. Now the magnetic moment becomes 7.5 $\mu_B$/Gd, in consistent with its valence state +2.5 after Li-implantation.

Thus, its quasi-1D ferromagnetism can be naturally attributed to a double-exchange-like process, although microscopically the dominant exchange here seems to be direct between Gd's spins. Here the localized 4$f$ moments play as the spin background while the itinerant 5$d$ electrons are the medium to connect ferromagnetism and metallicity, similar to the case in manganites ($t_{2g}$ spin background plus itinerant $e_g$ electrons) [38]. According to Fig. 4c, the effective magnetic coupling along the $a$-axis ($J_a$) should be much stronger than $J_b$ and $J_{ab}$ (as illustrated in Fig. 3d), which can be derived using the following equations and DFT energies:

$$E_{\text{FM}}=E_0+J_a+J_b+1/2J_{ab}, \quad E_{\text{S-AFM}}=E_0+J_a-J_b-1/2J_{ab},$$

$$E_{\text{Z-AFM}}=E_0-J_a-1/2J_{ab}, \quad E_{\text{DS-AFM}}=E_0+J_a+J_b-1/2J_{ab}, \quad (1)$$

where $E_0$ denotes the energy of the non-magnetic part. All DFT energies are recalculated based on the optimized lattice of FM state. Aforementioned argument is fully confirmed: $J_a$=-43.5 meV, $J_b$=-1.5 meV, and $J_{ab}$=-1.0 meV (for normalized spin |S|=1).

The MAE is also calculated, which prefers the out-of-plane easy axis (-0.2 meV lower than the in-plane one). Based on these coefficients, the magnetic transition can be simulated using the Heisenberg model and Monte Carlo method, as shown in Fig. 5a. Its FM $T_C$ is estimated to be 48 K, which is comparable to that of CrI$_3$ monolayer ($T_C$=45 K) [1] but its saturated magnetization is 2.5 times larger.

The anisotropic conductance is calculated by solving the semi-classical Boltzmann transport equations. As compared in Fig. 5b, the conductivity along the $a$-axis of FM state is metallic (i.e., decreasing with increasing temperature), but the perpendicular one is mostly due to the quantum tunneling and thus almost unaffected by temperature. The former one is much higher than the later, e.g. ~38 times at 6 K. Although this high anisotropy decreases with increasing temperature, it remains ~26 times at 30 K.

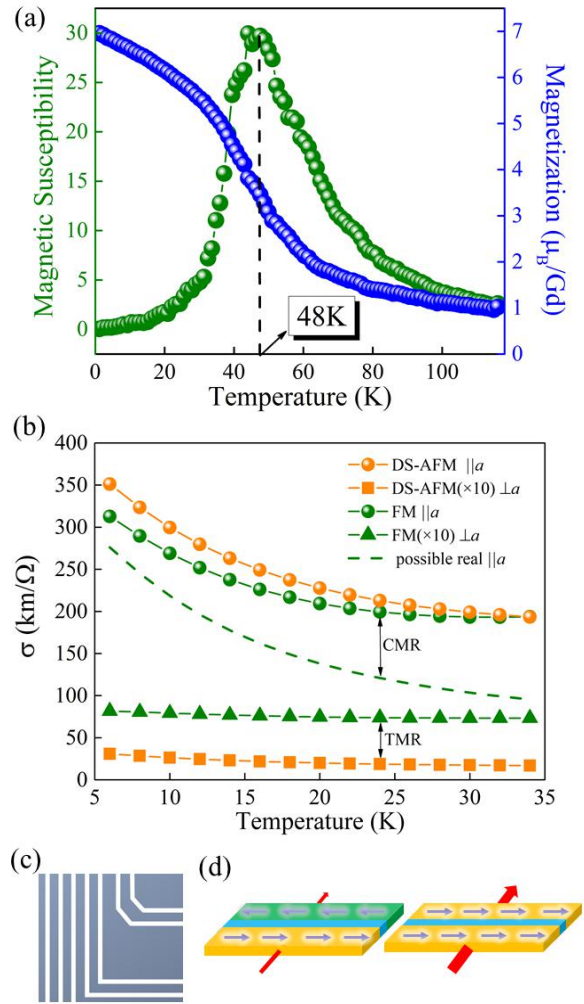

**Fig. 5.** Magnetic transition and anisotropic conductance of (GdCl$_3$)Li monolayer as a function of temperature. (a) Monte Carlo simulated magnetic susceptibility and magnetization. (b) Conductivities along different directions and with different magnetic orders (FM and DS-AFM). The broken curve denotes the possible real case with qualitatively correction of thermal fluctuation effects to the FM order. The CMR and TMR effects are indicated by arrows between curves. (c) Schematic of potential application as nano cabling. (d) Schematic of TMR between stripes. Red arrows: the tunneling currents.

Due to the double-exchange-like process, a moderate colossal magnetoresistivity (CMR) effect is naturally expected [as indicated in Fig. 5b], according to the experience from manganites [38]. In addition, the highly anisotropic transport behavior is also similar to that observed in strained manganites [39, 40], and can be utilized as the natural cabling in nanoelectronics with an ultra-narrow linewidth limit down to ~1 nm, as sketched in Fig. 5c.

Furthermore, since the DS-AFM state is energetically proximate to the FM one, its conductivity is also shown in Fig. 5(b). As expected, the conductivity along the $a$-axis is similar to the FM one, with slightly higher value (e.g., 12% higher at 6 K and 3% at 30 K). However, the quantum tunneling across the stripes is seriously suppressed in the DS-AFM state, e.g., ~38% of the FM one at 6 K and ~26% at 30 K. This effect is the quantum tunneling magnetoresistivity (TMR) effect, as sketched in Fig. 5(d).

## 4. Conclusion

In summary, we proposed an orthorhombic GdCl$_3$ monolayer with unique ion hendecahedra, which can be easily exfoliated from its existing vdW bulk. This structure is very rare in two-dimensional magnets, which is naturally isotropic. Despite its plain low-temperature antiferromagnetism due to the $4f^{\,7}$ electronic configuration, its ground state can be transformed into ferromagnetism upon electron doping, with a moderate Curie temperature comparable to CrI$_3$ but much larger saturated magnetization. More interestingly, the Li atom intercalation causes a significant structural reconstruction, leading to highly anisotropic conductance and magnetoresistivity, which may be uniquely useful in nanoelectronics.

## Credit author statement

S.D. and X.Y. conceived and supervised the project. H.Y. performed the DFT calculations. J.C. performed the MC simulation. J.J.Z. performed the transport calculation. N.D. and X.Z. provided support to calculations. H.Y., X.Y., and S.D. wrote the manuscript. All authors discussed results.

## Declaration of competing interest

The authors declare that they have no known competing financial interests or personal relationships that could have appeared to influence the work reported in this paper

## Acknowledgments

Work was supported by the National Natural Science Foundation of China (Grant No. 11834002). J.C. was supported by Postgraduate Research & Practice Innovation Program of Jiangsu Province (No. KYCX21_0079). We thank the Tianhe-II of National Supercomputer Center in Guangzhou (NSCC-GZ) and the Big Data Center of Southeast University for providing the facility support on the numerical calculations.

## Appendix A. Supplementary materials

Supplementary data to this article can be found online at https://doi.org/10.1016/j.mtpphys.2022.xxxxxx.